\documentclass[usenatbib]{mn2e}
\usepackage{epsfig}
\usepackage{amsmath}
\usepackage{amssymb}
\def\gtrsim{\mathrel{\raise0.35ex\hbox{$\scriptstyle >$}\kern-0.6em
\lower0.40ex\hbox{{$\scriptstyle \sim$}}}}
\def\lesssim{\mathrel{\raise0.35ex\hbox{$\scriptstyle <$}\kern-0.6em
\lower0.40ex\hbox{{$\scriptstyle \sim$}}}}

\def\h50{h_{50}}

\title[Line emission in BCGs]{Line Emission in the Brightest Cluster Galaxies of the NOAO Fundamental Plane and Sloan Digital Sky Surveys}

\author [Edwards et al.]
                   {Louise~O.V.~Edwards$^1$, Michael J. Hudson$^2$, Michael L. Balogh$^2$, Russell J. Smith$^3$ \\
$^1$D\'{e}partement de Physique, G\'{e}nie Physique et d'Optique, Universit\'{e} Laval and Observatoire du mont M\'{e}gantic, Qu\'{e}bec, QC, \\      ~G1K 7P4, Canada\\
$^2$Department of Physics and Astronomy, University of Waterloo,
  Waterloo, ON, N2L 3G1, Canada\\
$^3$Department of Physics, Durham University, Durham UK,DH1 3LE\\
}

\date{\today}

\begin{document}
\maketitle

\begin{abstract}

  We examine the optical emission line properties of Brightest Cluster
  Galaxies (BCGs) selected from two large, homogeneous datasets. The
  first is the X-ray selected National Optical Astronomy Observatory
  Fundamental Plane Survey (NFPS), and the second is the C4 catalogue
  of optically selected clusters built from the Sloan Digital Sky
  Survey Data Release ~3 (SDSS DR3). Our goal is to better understand
  the optical line emission in BCGs with respect to properties of the
  galaxy and the host cluster.  Throughout the analysis we compare the
  line emission of the BCGs to that of a control sample made of the
  other bright galaxies near the cluster centre. Overall, both the
  NFPS and SDSS show a modest fraction of BCGs with emission lines
  ($\sim 15$\%). No trend in the fraction of emitting BCGs as a
  function of galaxy mass or cluster velocity dispersion is found.
  However we find that, for those BCGs found in cooling flow clusters,
  71\ensuremath$^{+9}_{-14}$ have optical emission.  Furthermore,
  if we consider only BCGs within 50kpc of the X-ray centre of a
  cooling flow cluster, the emission-line fraction rises further to
  100\ensuremath$^{+0}_{-15}$~\%. Excluding the cooling flow
  clusters, only \ensuremath{\sim} ~10~\% of BCGs are line emitting,
  comparable to the control sample of galaxies.  We show that the physical origin of the emission line activity  varies: in some cases it has LINER-like line ratios, whereas in others it is a composite of star-formation and LINER-like activity.  We conclude that the presence of emission lines in BCGs is directly related to the  cooling of X-ray gas at the cluster centre.
\end{abstract}


\section{Introduction}

The brightest cluster galaxy (BCG) is typically a giant, red
elliptical or cD galaxy, located near the centre of the gravitational
potential.  It is likely that a rich history of galaxy-galaxy
interactions and mergers is responsible for the unique morphology of
such galaxies.  This is supported indirectly by several pieces of
evidence, including: the luminosity of the BCG is correlated
with the cluster mass \citep{sco57,lin04} and X-ray luminosity
\citep{Hud97}; BCGs in the most X-ray luminous clusters are larger and have surface
brightness profiles which are less steep than their low X-ray
luminosity counterparts \citep{Bro05}; and the
velocity dispersion of BCGs rises less steeply with luminosity than
for other bright galaxies \citep{von06}.
All this indirect evidence for a rich merger history is supported by
high resolution imaging of BCGs obtained with the {\it Hubble Space
  Telescope}, which has revealed that the cores of these galaxies can
be complex, often showing multiple nuclei and prominent dust
signatures \citep{lai03}. 

However, the evolutionary history of these galaxies is still not
completely understood, and work on cooling flow (CF) clusters has hinted at another possible mechanism for adding to the stellar mass of
BCG.  The original cooling flow hypothesis is that hot cluster X-ray
gas cools and condenses out of the intracluster medium (ICM) into the
cluster's potential well, forming molecular clouds and stars \citep{fab94}.
This drop-out occurs at the centre of the cooling flow, within the
cooling radius, ie. onto the BCG. Cooling flow clusters are common in
the local universe \citep[making up about 50\% of the population in an X-ray
flux limited sample][]{per98,che07}), and cD
galaxies are often found at the centre of these systems. Because of
this, a link between the cooling X-ray gas and recent star formation
in the BCG has been discussed for many years \citep{fab94}. Convincing
observations which support this idea have been presented: a blue and
UV-colour excess \citep{mcn96,mcn04,hic05}, molecular gas
\citep{edg02,jaf01,sal03} and H$\alpha$ emission
\citep{cra99,don00,cra05} have all been seen in the cooling flow
cluster's BCG. However, although the morphology of the H$\alpha$
emission is diffuse and filamentary, indicating star formation in some
CF BCGs, in others it is very compact and more characteristic of AGN
dominated emission \citep{don00,edw07,hat07}. As well, \citet{von06}
recently showed that optical emission lines in BCGs predominantly
arise from LINER emission, rather than normal star formation.
  
More recent X-ray satellite measurements from { \it Chandra } and {
  \it XMM} have shown that the gas does not cool directly from the hot
X-ray phase through to the cool molecular gas phase \citep{boh02}, but
rather that a large amount of the cooling gas is being reheated before
condensing out of the ICM. The current paradigm is that AGN activity
in the BCG is reheating the cooling X-ray gas, which implies a more
complicated feedback process between the cooling gas and the central
galaxy \citep[e.g][]{piz05}. This leads to revised, predicted mass
deposition rates that are now in reasonable agreement with the
observed values \citep{boh02,piz05}. The observed molecular and ionic
gas may be attributed to a small amount that has cooled from the
cooling flow; alternatively, the H$\alpha$ may be excited by the AGN
itself. Detailed studies of star formation indicators in galaxy groups
and clusters in cooling flows and non-cooling flows, discriminating
between those with and without AGN activity, are required in order to
analyze the relative importance of the different processes.

A correlation between optical line emission in the BCG and cluster
properties has been explored by several authors, most notably
\citet{cra99}, who found that 27\% of BCGs have optical line emission,
and that the projected distance from the BCG to the X-ray centre is
less for line emitting galaxies than for non-emitting galaxies.
Recently, \citet{von06} and \citet{bes06} have explored the properties
of BCGs in the SDSS, using the C4 cluster catalogue \citep{mil05}.
These authors find that radio-loud AGN activity is more frequent in
BCGs than in other galaxies of the same mass, but that this frequency
does not depend strongly on cluster velocity dispersion. On the other
hand, in their study of  radio-loud properties of an X-ray selected
cluster sample, \citet{lin06} find the overall radio-loud fraction
to be 30\% in BCGs, and that the fraction is higher in more massive
clusters.  Importantly, 
\citet{von06} find that many of these radio-loud galaxies would not necessarily
be identified as AGN from their optical emission lines and, in fact,
that optical AGN activity appears to be {\it less} frequent among BCGs
than other cluster galaxies of similar mass.  The interpretation is
complicated by the fact that the radio-selected galaxy sample, though
restricted to red galaxies, could be contaminated by galaxies in which
the low luminosity radio emission arises from star formation, rather
than AGN activity.
  
In this paper, we explore optical line emission in BCGs with respect
to properties of the galaxy and the host cluster, using two large,
homogeneous datasets. One sample is taken from the X-ray selected,
National Optical Astronomy Observatory Fundamental Plane Survey
(NFPS), for which the X-ray properties are known for all clusters. 
For many of these clusters, we are able to identify those with
short cooling times (CF clusters) based on {\it ROSAT}, {\it
  Chandra}, or {\it XMM-Newton} observations.  We complement this sample with
optically-selected clusters drawn from the Sloan Digital Sky Survey
Data Release 3 (SDSS DR3), which is not biased toward X-ray luminous
clusters, and is therefore more representative of the cluster
population.  In addition, the greater spectral coverage of the SDSS
allows us to use emission line ratios to identify whether the emission
arises predominantly from composite HII region and LINER activity, or from LINER activity alone
(pure HII-region, and Seyfert-like emission are both rare).  

The paper is organized as follows. In section \ref{data}, we introduce
our galaxy samples and selection criteria. In section \ref{results},
we report our results. For the X-ray selected NFPS we compare the
frequency of H$\beta$ emission in BCGs as a function of BCG magnitude
and distance to the cluster centre.  Similar results are found for the
SDSS sample, based on the H$\alpha$ emission line. However, we find
that there are differences between the BCGs in the two samples, which
we can attribute to the nature of the X-ray emitting gas.  In section
\ref{discussion}, we consider the impact of our results on various
galaxy and cluster formation hypotheses, and summarize our results. We
conclude in section \ref{conclusion}. Unless otherwise stated our
analysis assumes the values $\Omega_{\mathrm m}$=0.3, for the matter density parameter,  $\Omega_{\Lambda}$=0.7 for the cosmological constant, and $H_0$=100 km/s/Mpc for the Hubble parameter.  L$_{X}$ refers to the bolometric X-ray luminosity
throughout. 

\section{Data and Sample Selection}\label{data}

For both NFPS and SDSS, our goal is to identify the BCG and a similar
sample of luminous ``control'' galaxies that is located in the
inner regions of the cluster.  We first discuss the sample selection
of the NFPS. 

\subsection{NOAO Fundamental Plane Survey}\label{nfp}

The NFPS is an all sky study of 93 X-ray selected rich clusters with
redshifts between 0.010 and 0.067 \citep{smi04}. The goals of the
project are to measure cosmic flows on the scales of 100 h$^{-1}$ Mpc
and to build a large homogeneous sample with which to investigate
physical effects and environmental influences on early-type galaxy
evolution and formation \citep{smi04}. The spectroscopic observations are made though a fiber diameter of 2" and are limited to red sequence galaxies; galaxies more than 0.2
magnitudes bluer than the red sequence were not generally observed
spectroscopically. There are spectra for 5388 galaxies with a
wavelength coverage between 4000 and 6100~\AA~ and a resolution of
\ensuremath{\sim} 3 \AA ~\citep{smi04}. 

\subsubsection{Cluster, BCG, and Control Sample Definitions}\label{nfpss}

In order to respect the completeness of the NFPS cluster sample, we
exclude the 13 clusters that were observed serendipitously and that did
not meet the original $L_{X}$ limit of $10^{42.6}$ erg/s.  For each cluster, the
redshifts were used to calculate the cluster velocity dispersion and the radius at which the mean density interior to the cluster is 200 times the critical density ($\sigma_{cl}$ and r$_{200}$, respectively). We used the prescription $r_{200}$=$\sqrt{3}$$\sigma_{cl}$/1000~km/s/Mpc as derived in \citet{car97}.  Note that r$_{200}$ is typically of order 1.5 $h^{-1}$ Mpc, much larger than
the cooling radius of a typical cooling-flow cluster (about 200 kpc).  The
centre of the cluster is taken to be at the peak of the X-ray emission
\citep{ebe96,ebe00}.  The galaxies are then assigned to the clusters based on a
radial and a velocity weighting as in \citet{smi04}.  For this
analysis, we are interested in only the bright galaxies in the central
regions of the cluster and so include a magnitude limit of
$M_{K}<-24$, based on K-band total magnitudes obtained from 2MASS
catalogues \citep{skr06}.   This is about half a magnitude brighter
than the characteristic K-band magnitude of $M_{K*} = -23.55$
\citep[][for $H_0 = 100$ km/s/Mpc]{lin04b}.  Since we are interested only in galaxies occupying a similar environment as the BCG, ie. in the central regions of the cluster, we consider only galaxies within $0.5r_{200}$, and with velocity differences with respect to the cluster mean velocity less than twice the cluster velocity dispersion. 

The BCG is then defined as the first rank cluster galaxy using the
K-band magnitudes.  In twenty cases, the BCG was not observed
spectroscopically due to constraints in the fiber positioning, and these clusters have been excluded from our analysis. Our final sample consists of 60 clusters with BCGs, as summarized in Table \ref{restab} and listed in Table \ref{bcgtab}. Our
control sample consists of the 159 other bright ($M_{K}<-24$) galaxies
within the same radial and velocity cuts described above. The BCGs
are, of course, excluded from the control sample. 

For the NFPS, we use the stellar absorption-corrected H$\beta$
emission as an indicator of star formation activity, since H$\alpha$
is not generally available in these spectra.  Although $H\beta$
emission is relatively weak, the high signal-to-noise ratio of these
spectra allow us to measure its strength reliably after correcting for
underlying stellar absorption (see the errors quoted in Table
\ref{bcgtab}). The observed galaxy spectra are divided by best fit
absorption template stellar population synthesis models from
\citet{vaz99}, which have been redshifted and broadened to match the
velocity dispersion of the observed galaxy. Subsequently, the H$\beta$
equivalent width is measured directly, without assuming a particular
line profile, from the ratio of the observed spectrum to the best fit model \citep{nel05}. A thorough discussion of the emission line measurements can be found in \citet{nel05}. For two of the NFPS BCGs, in Abell 780 and in Abell 1795, the nebular emission is strong enough that our standard methods for obtaining 
reliable velocity dispersions, and hence stellar absorption corrections, fail.  For galaxies of similar magnitude, the typical stellar absorption at H$\beta$ is $\sim$1.5~\AA~ with an uncertainty of 0.3~\AA. This error dominates our uncertainity in the total equivalent width of emission for these two special cases. For the BCGs in Abell 780 and Abell 1795, the H$\beta$ equivalent widths are 7.8 \AA~ and 7.2 \AA, respectively. However, to achieve a plot that is more easily read, these points are set to respective lower limits of 3.3~\AA~and 3.7~\AA~in Fig. \ref{relkcont} and Fig. \ref{hbrcont}. 

We will define emission-line galaxies to be those with an equivalent width $>$ 0.5 \AA. Since the H$\alpha$ and [N{\sc ii}] lines are generally unavailable, we are unable to use \citet[hereafter BPT]{bpt81} diagrams to reliably distinguish emission due to star formation from that arising from AGN activity. 

\subsubsection{Cooling Flow definition}
In general, we designate an NFPS cluster as ``cooling flow'' (CF) or ``non-cooling
flow'' using mass deposition rates and X-ray cooling times from
published catalogues \citep{per98,whi00,all01,bir04}. This is a
somewhat subjective classification complicated by the fact that not only are mass
deposition rates calculated from {\it ROSAT} observations 
typically 2-10 times higher than those calculated from {\it Chandra}
observations \citep{boh02}, but also higher resolution spectra from {\it XMM Newton} are not well matched to an isobaric cooling model \citep{pet03}; thus the mass deposition rate may not be an exact indicator of a cooling flow cluster. Therefore, whenever possible, we prefer to use recent cooling flow designations based on the presence of a central temperature gradient in {\it XMM Newton} or {\it Chandra} observations. For the rest, we are left with using the mass deposition rate based on {\it ROSAT} data as an indicator. For the 11 cases where observations are from the {\it Chandra} or {\it XMM Newton} satellites, we define a CF cluster to be one
with a mass deposition rate $\dot{M} > 0$; otherwise, we require
$\dot{M}>100 M_{\sun}\mbox{yr}^{-1}$. Within this framework we have 14 CF
clusters and 19 non-CF clusters.  The CF status of another 27 clusters
is unknown. Clearly this is not an unassailable definition; however,
it is likely that most, if not all, of the clusters we classify as CF
clusters really do have short cooling times in the centre.  Some
galaxies with low mass deposition rates will undoubtedly fall in our
non-CF sample.  With this in mind, our results are not sensitive to
this definition. We further discuss this point in Section
\ref{sec-cfdef}, as well as a continuous method of defining a CF
cluster, based on an excess of observed X-ray luminosity to predicted
values \citep{mcc04}. 
\subsection{Data from the Sloan Digital Sky Survey}\label{sec-data}

Our second sample of galaxy clusters is derived from the C4 catalogue
\citep{mil05}, based on the third data release (DR3) of the SDSS. 
This release covers 5282 square degrees in imaging, and 4188 square
degrees in spectroscopy \citep{aba05}. Imaging was taken in five
optical bands, u', g', r', i' and z' with a median spatial resolution
of 1.4" in r'. Spectra are observed with a  aperture diameter of 3",
cover the wavelength range from 3800 to 9200 \AA~ and have a spectral
resolution of $\sim$3~\AA~ in r'. Redshifts measured from these
spectra are accurate to $\sim$ 30~km/s \citep{aba05}.  The H$\alpha$
line strengths are measured by fitting gaussians to the line profile
in a standard pipeline \citep{sto02}. For $W_\circ(H\alpha)>5$\AA, the
equivalent width uncertainty is less than 20\%.   For weaker lines, $W_\circ(H\alpha)<5$\AA~
the errors are known to be large \citep{gom03}.  We have made no
correction to the emission-line strengths for underlying stellar
absorption.  For our purposes this is safe to neglect because, whereas
even a modest star formation rate generates considerable H$\alpha$ emission,
the stellar absorption doesn't vary by more than $\sim1$~\AA~for
moderately old populations.   

As described in \citet{mil05}, clusters and groups are identified as overdensities in the multi-dimensional space spanned by position,
redshift, and five-colour photometry. There are 1106 clusters
identified in the SDSS DR3 using this algorithm. We further select
objects with redshifts $z<$ 0.10, to minimize the incompleteness of the sample. 
In order to reduce the number of clusters with uncertain velocity
dispersions, we exclude clusters flagged as having significant 
substructure. Specifically we include only clusters flagged as SUB=0, ie. those for which the ratio of the standard deviation of the cluster velocity dispersion profile to the mean cluster velocity dispersion is less  than 15 \citep{mil05}. This reduces the number of clusters in our sample to 825. 

\subsubsection{Cluster, BCG, and Control Definitions}\label{selectioncriteria}

For the SDSS we define the BCG and clusters in as close a manner as
possible to the NFPS case. Again, the BCG is the brightest galaxy in
the K band (and with $M_{K} <-24$), within half of r$_{200}$ of
the geometric centre of the cluster, and within two times the cluster
velocity dispersion. However, the geometric centres we use are
different than those in the C4 catalogue. We start with the C4 catalogue geometric centres measured using the luminosity weighted position average of all the galaxies within 1Mpc and four times the velocity dispersion. As this definition encompasses such a large area, multiple substructures along the line of sight heavily influence the position of the centre. For example, the geometric centre will be placed in between two obvious subclumps. Thus, we iteratively recalculate a luminosity-weighted centre using only galaxies within two times the cluster velocity dispersion, and with magnitude $M_{K} <-24$.   

Due to the incomplete spectral coverage of the central galaxies, there
are 154 cases in which the brightest cluster galaxy had not been
observed spectroscopically. We exclude these clusters from the sample. 
To ensure this is not introducing an important bias, we verified that
the subset of BCGs observed in the photometric catalogue but not in
the spectroscopic catalogue have an equivalent ($u'-r'$) colour
distribution to those in our final, spectroscopic sample. Finally, we
also remove those clusters whose geometric centres are within 15' of
a survey boundary, as well as those with measured
velocity dispersions greater than 1200 km/s.  The latter restriction
is made because such high values are usually a result of significant
contamination from line-of-sight substructure.  These selections leave
us with a final sample of 328 BCGs. 

Our control sample is built from the other bright galaxies near the
centre of the clusters, as in the NFPS case. In order to increase the
size of our control sample, we include clusters where the brightest
central galaxy was not measured spectroscopically. There are 526
control galaxies in 353 groups and clusters with a velocity dispersion
less than 1200km/s. We note that the number of control galaxies per
cluster is significantly larger in the X-ray luminosity 
selected NFPS clusters, probably because they are richer than the
optically-selected SDSS clusters. 

\begin{figure}
\centering
\epsfysize=3in
\epsfbox{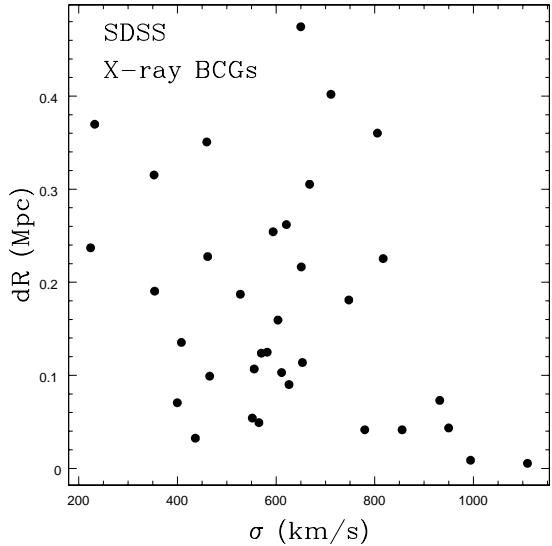}
\caption[Difference in X-ray and Geometric Centres]{
Cluster velocity dispersion as a function of the difference in the X-ray and geometric
  centres for the 35 SDSS BCGs that have available X-ray positions. The clusters with large
  discrepancies between the two centres are generally found to have
  geometric centres highly influenced by in-falling groups. 
  \label{drsig}}
\end{figure}

Because the centres of the NFPS clusters are based on the X-ray
centroid, whereas in the SDSS the geometric centre is used, it is
important to compare the two definitions. In order to find X-ray centers for the SDSS clusters, we have matched them to X-ray cluster catalogues including: NORAS \citep{boh00},
REFLEX \citep{boh01}, BAX \citep{sad04}, XBACS \citep{ebe96}, RASSCALS
\citep{mah99}, as well as \citet{pop04,mul98} and \citet{hor01}. Most of the
cluster catalogues are based on {\it ROSAT} observations, where the
flux limit is high.  This restricts our sample of SDSS clusters with
X-ray detections to the small subset of massive clusters at z ~$<$
0.03. There are 35 X-ray cluster matches, ie. cases where the X-ray
centre is within half of r$_{200}$ of one of our 328 SDSS clusters.  Fig. \ref{drsig} shows that for most of the matched clusters, there is good agreement between the X-ray and geometric centres. However, also depicted in the figure is how some cases exhibit differences of up to 0.5 Mpc, especially for lower mass clusters. It is unlikely that this is caused by an uncertainty in
the X-ray centres, as the centres of the NORAS, BCS, and XBACS samples were found by
determining the two-dimensional centre of mass using the Voronoi
Tessellation and Percolation method \citep{ebe93}, and are generally
accurate to about 1' \citep{ebe98}, corresponding to $\sim 90$ kpc at
$z\sim 0.08$.  More likely, is that the geometric centers do not trace the gravitation potential of the
cluster as well as the X-ray centres. We note that for many of the cases
in which the centres are discrepant by $>200$ kpc,
the geometric centre appears to be contaminated by in-falling groups; which is not surprising as it is
measured using a typical line of sight projection of $\sim 3$ Mpc
(for $\sigma=500$km/s).  

\subsubsection{Emission Line diagnostics}\label{eline_sdss}

\begin{figure}
\centering
\epsfysize=3in
\epsfbox{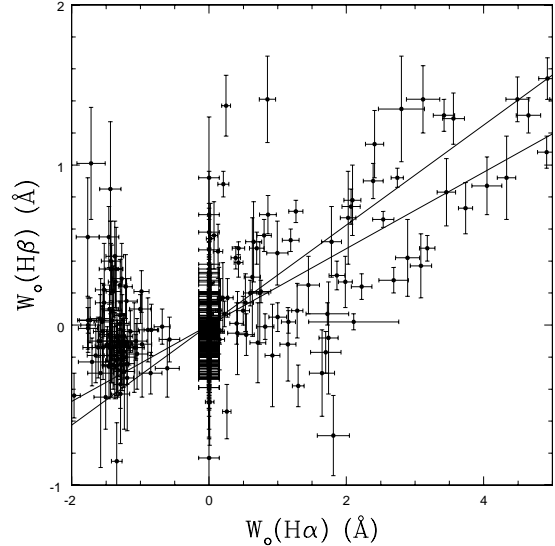}
\caption[Relationship of H$\alpha$ and H$\beta$ W$_{o}$]{
  The absorption-corrected NFPS H$\beta$ equivalent width versus the SDSS H$\alpha$ equivalent
  width (uncorrected for absorption) for galaxies present in both surveys. The points with
  W$_{o}$(H$\alpha$)~$= 0$ are non-detections (plotted with the average H$\alpha$ error of 0.15\AA), and those with
  W$_{o}$(H$\alpha$) $< 0$ are detected in absorption. We identify
  line-emitting NFPS galaxies as those with
  W$_{o}$ (H$\beta$) $\ge$ 0.5~\AA. Correspondingly, we use 
  a value of W$_{o}$ (H$\alpha$) $\ge$ 2~\AA\ for the SDSS galaxies. The best fit lines are
  constrained to go through (0,0) and are only fit to galaxies with
  W$_{o}$ (H$\alpha$) $\ge$ 2~\AA~ and ignore an outlier at W$_{o}$ (H$\alpha$) $\approx$ 40.  The two different lines are fitting for
  W$_{o}$ (H$\beta$) as a function of W$_{o}$~(H$\alpha$) or vice
  versa. \label{cfewzoom}}  
\end{figure}

Optical emission line galaxies in the SDSS are identified from the
H$\alpha$ line, which is the line in our wavelength range that is most
sensitive to star formation activity.  To ensure a fair comparison
with the NFPS, we must choose a threshold in H$\alpha$ equivalent
width that is comparable to the H$\beta$ limit used in that
survey.  Recall that the lines are measured using different
techniques, from spectra of different
resolution and signal-to-noise, and obtained with different fibres;
furthermore, the NFPS H$\beta$ measurements are corrected for underlying
stellar absorption, while the SDSS H$\alpha$ lines are not.  Therefore
we opt for an empirical ``calibration'' between the two, by plotting
the NFPS H$\beta$ equivalent widths
against the SDSS H$\alpha$ equivalent widths, for galaxies that appear
in both surveys (Fig. \ref{cfewzoom}).  Although there is plenty of
scatter, there is a strong correlation between the two lines, and we
find the H$\alpha$ index is about four times larger than the H$\beta$;
encouragingly, this is comparable to 
 the factor of 4.5 derived by comparing
 the H$\alpha$/H$\beta$ ratio for the subset galaxies in the NFPS for
 which measurements of both emission lines are available
 \citep{nel05}.  Thus, an H$\alpha$
equivalent width cut of 2~\AA\ is comparable 
to an H$\beta$ equivalent width of 0.5 \AA, and the fraction of
galaxies in our samples above either of these thresholds is similar.  Note that the
points at $W_\circ(H\alpha)=0$ are non-detections  (plotted with the average uncertainty of 0.15\AA), and those with
$W_\circ(H\alpha< 0)$ are detected in absorption.   We have
experimented with SDSS H$\alpha$ equivalent width cuts of
2\ensuremath{\pm}1~\AA~and find that our final results do not change
significantly. The good correlation also gives us additional confidence in the
template correction used to correct the H$\beta$ equivalent
widths for stellar absorption, which is relatively much more important
here than for H$\alpha$ (the average stellar absorption strength
is $\sim1$~\AA, with variations at the
$\sim0.15$~\AA~level).

Due to the proximity of the galaxies in our
sample and the finite fiber size though which they are observed, the 
 amount of extended line emission could be underestimated. However,
 as we do not know if the emission is extended we have not explicitly
 accounted for the finite fiber size, nor for the difference in fiber
 diameters between two surveys. Here again, we argue that the effect
 on the line strength measurements is calibrated by our use of an
 empirical relation.   

H$\alpha$ line emission may arise from either ionization by hot stars, or 
ionization from AGN activity.  We use the AGN classification taken
from the emission-line analysis discussed in \citet{kau03}, which is
based on H$\alpha$/[N{\sc ii}] and [O{\sc iii}]/H$\beta$
diagnostic ratios, which separates AGN, star-forming, and composite (intermediate between star-forming and AGN) regions on BPT diagrams.  Table \ref{restab} shows that, in our SDSS sample, $\sim 65$ \% of
galaxies with emission have line ratios consistent with
AGN or composite emission, and this fraction may be somewhat higher for
the BCG population, relative to the control galaxies. Using \citet{ho97} to separate Seyferts from LINERs, along with the \citet{kau03} definitions, we find that~$\sim33$\% of emitting SDSS BCGs can be reliably measured as LINERs,~$\sim27$\% as composite;
Seyfert-like emission is negligible (~$\sim3$\%).  If we assume the ionizing source for the HII regions 
is stellar, and that all of the H$\alpha$ emission is within the fibre diameter, then typical H$\alpha$ luminosities of our emitting BCGs correspond to star formation rates of $\sim0.5$ to $\sim1.6$~$M_{\sun}
\mbox{yr}^{-1}$ \citep{ken98}. 

\subsubsection{Galaxy Colour}

\begin{figure}
\epsfysize=3in
\epsfbox{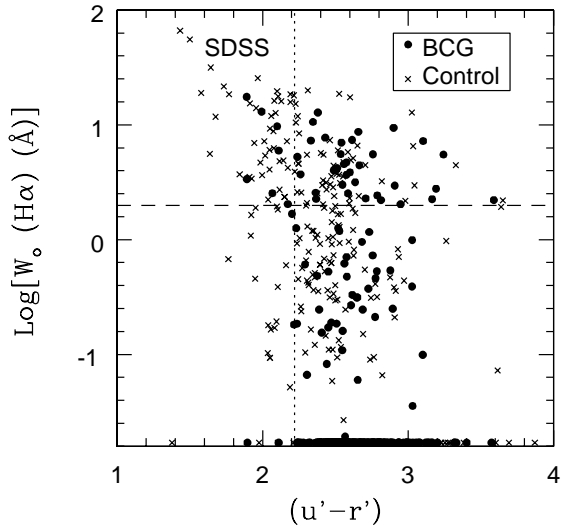}
\caption{The logarithm of W$_{o}$ (H$\alpha$) as a function of ($u'-r'$)
  colour of the galaxies in the SDSS. The crosses are the control sample,
  and the filled circles are the BCGs.  The colour break of 2.2 is
  shown as the vertical dotted line, and the horizontal dashed line
  separates the emitting galaxies from the quiescent. The galaxies at 
  Log[W$_{o}$ (H$\alpha$)]  = -1.7 are those with H$\alpha$
  absorption.\label{hacolb}}
\end{figure}

Fig. \ref{hacolb} shows the distribution of H$\alpha$ equivalent width
as a function of ($u'-r'$) colour.  Following \citet{str01}, we use
the value of 2.2 for the ($u'-r'$) colour separation of blue and red
galaxies. As the NFPS galaxies are red selected, we include a
corresponding colour cut of ($u'-r'$)~$>$~2.2 for the SDSS galaxies
when directly comparing to the NFPS results (as in Section
\ref{nfpresults}). This colour cut excludes 10 of the 328 SDSS BCGs (only one with X-ray data) as well as 53 control galaxies. 

As seen in Fig. \ref{hacolb}, most BCGs are redder than this cut
whether or not they show optical emission. Thus, if the line emission
is due to star formation, it does not dominate the global colours of
these BCGs, which are presumably quite old.  Because of this, results
based on BCGs in the NFPS are not likely to be affected by the colour
cut in that sample.  For the control sample, most of the emission-line
galaxies are bluer than our cut. Therefore we expect the fraction of
control galaxies with emission, as presented in Section~\ref{results},
to be biased low relative to the BCGs.  This is evident from
Table~\ref{restab}: in the red-selected SDSS sample, the overall
fraction of emission line galaxies is similar ($\sim ~11$ \%) for both
control galaxies and BCGs, while for the unrestricted sample the
fraction of control galaxies with emission is $\sim ~18$\%.  This
difference is not large enough to affect any of our conclusions in
Section~\ref{results}, however. 

\section{Results}\label{results}

\label{nfpresults}

\begin{table*}
  \centering
   \caption{
     Summary of Results. The first column presents the survey studied, NFPS or SDSS. SDSS (RED) includes only the red-selected SDSS galaxies, and SDSS (X-ray) only those with X-ray counterparts. Column 2 indicates if the results are for the BCGs or Controls. Column 3 shows the total number of galaxies for each sample. The number of these which are emitting is shown in column 4, and column 5 shows the fraction of emitting galaxies. Column 6 lists the number of strongly emitting (W$_{o}$(H$\alpha$)$>2$\AA) galaxies in each sample for which the emission line ratios are characteristic of AGN activity (usually LINER). We present the number of galaxies in our sample which are known to belong to a CF cluster in column 7. Column 8  shows the number of these galaxies in CF which are also emitting, and the final column shows the fraction of emitting galaxies in CF clusters.} 
 \label{restab}
    \begin{tabular}{llccccccc} \hline \hline
Survey & Sample & Total &Emitting &Emitting  &AGN or Comp. & Known&Emitting&Emitting\\
              &                &          &        &      Fraction(\%) &   W$_{o}$(H$\alpha$)$>2$\AA &CF   &CF&Fraction(\%)\\
\hline

NFPS & BCGs    &  60 & 12 & 20\ensuremath{\pm}6 & N/A & 14 & 10 &71\ensuremath$^{+9}_{-14}$\\ 
            & Controls&159 & 15 &9\ensuremath{\pm}2& N/A& 36 & 5 &14\ensuremath{\pm}6\\
SDSS & BCGs   &  328 & 42  &13\ensuremath{\pm}2& 31 & N/A & N/A&N/A\\ 
           & Controls& 526 & 94  & 18\ensuremath{\pm}2&57 &N/A & N/A&N/A\\
SDSS (RED)& BCGs & 318 & 35 &11\ensuremath{\pm}2& 25 &N/A & N/A &N/A\\
    & Controls& 446 & 51 & 11\ensuremath{\pm}2& 39 & N/A & N/A&N/A\\
SDSS (X-ray)& BCGs &34&6&18\ensuremath$^{+8}_{-5}$&3&N/A&N/A&N/A\\
\hline
\end{tabular}
\end{table*}

In this section we examine the line emission in the galaxies in our
cluster samples as a function of X-ray luminosity, K-band
magnitude, and distance from the centre of the cluster.  The numbers
of emission-line galaxies in both surveys are given in column 4 of
Table \ref{restab}. There is a higher fraction of emitting galaxies
among the NFPS BCGs; 20\ensuremath{\pm}6\% are emitting, compared to
only 9\ensuremath{\pm}2\% of the controls.  If we look only at those
BCGs identified with a CF cluster we find an even higher fraction of
them show emission, 71\ensuremath$^{+9}_{-14}$, while the control
sample is unchanged (14\ensuremath{\pm}6\%).  On the other hand, only
11\ensuremath{\pm}2\% of the red BCGs selected from the SDSS sample
show emission, comparable to that of the control population
(11\ensuremath{\pm}2\%). In this section we will explore these trends, and differences between
the two samples, in more detail. We use errors derived from the posterior probability distribution where sample sizes are small.

\subsection{Dependence on the Presence of a Cooling Flow}\label{sec-cfdef}

\begin{figure}
\centering
\epsfysize=3in
\epsfbox{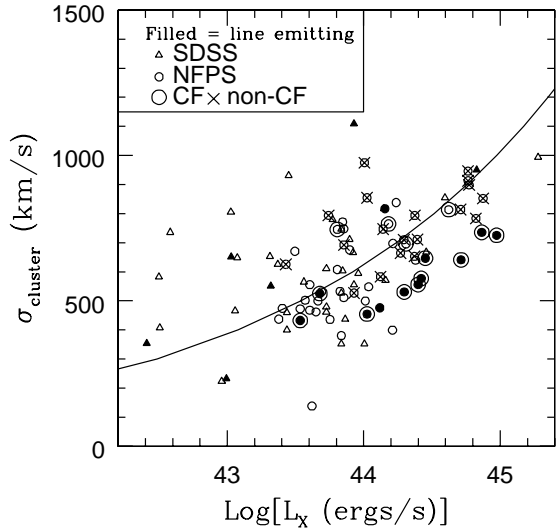}
\caption[Cluster Velocity Dispersion as a function L$_{X}$]{The cluster
  velocity dispersion as a function of the bolometric X-ray
  luminosity. The solid line is the preheated model from
  \citet{bab02}, which is known to provide a good match to non-CF
  clusters \citep{mcc04,bal06}.  Known CF clusters in
  our sample do generally lie to the right of this model.  \label{withmodel}} 
\end{figure}

The most prominent result from the NFPS clusters is that the presence
of a cooling flow is highly correlated with the presence of emission
lines in the BCG. Fig. \ref{withmodel} shows the bolometric X-ray
luminosity against the cluster velocity dispersion (a proxy for
dynamical mass). We can see that most of the cooling flow clusters
have larger X-ray luminosities for their mass, and that it is these clusters that have
a BCG with line emission. Therefore, in the rest of the NFPS analysis
we separate our sample into CF and non-CF subsets.  Notice that in
Fig.~\ref{hbkcont} and Fig.~\ref{hbrcont} (which we discuss below),
none of the emitting BCGs are non-CFs.  As mentioned in Section
\ref{nfpss}, we use a rather strict cut for our definition of a CF, requiring {\it ROSAT} based $\dot{M}>100 M_{\sun}{yr}^{-1}$. However, changing our
arbitrary definition of a CF cluster (e.g. to those with $\dot{M}> 10
M_{\sun}\mbox{yr}^{-1}$) does not significantly change our results. 

It would be useful to have a way to identify likely cooling-flow
clusters without the need for high quality surface brightness and
temperature maps.  Recently, \citet{mcc04} has shown that CF clusters
show significantly higher total X-ray luminosities, relative to their
total dynamical mass, consistent with predictions from steady-state
cooling models.  On the other hand, non-CF clusters can be well
modelled with cosmological haloes in which the gas has been preheated
to $\sim 300-500$ keV cm$^2$ \citep[see also]{bal06}.  This suggests
that one could use the excess X-ray luminosity relative to these
preheated models as an indicator of CF status. The solid line in Fig. 
\ref{withmodel} represents models for clusters with preheating from
\citet{bab02}.  Indeed, using this method arrives at results for the
emission fraction which are very similar to those obtained when we use
the mass deposition rate to define CF clusters.  The separation is not
as striking as in \citet{mcc04}, probably because the measured
velocity dispersion can be systematically affected by substructures. 

\subsection{Magnitude Dependence}

\begin{figure*}
\centering
\epsfysize=9cm
\epsfbox{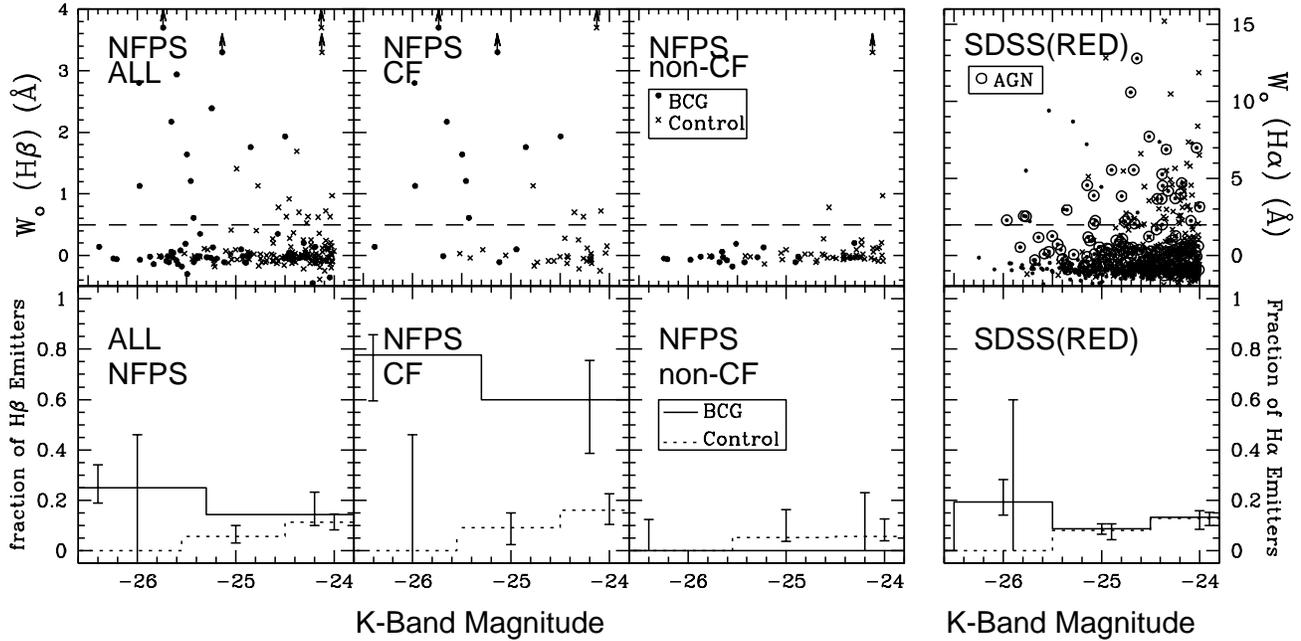}
\caption[Emission line strength as a function of K magnitude]{Line
  Emission as a function of K-Band Magnitude. \underline{Top  Panel}:
  The equivalent width of the line emission as a function of K-band
  magnitude for the NFPS galaxies on the left, and for the SDSS
  galaxies on the right. The NFPS galaxies are in subsets: a) All
  galaxies, including those in clusters where the CF status is unknown,
  b) those which are in known CF clusters, and c) those which are in
  known non-CF clusters. Filled circles represent BCGs, Xs for control
  galaxies, and open circles indicate LINER emission (for the SDSS
  sample only). The dashed line represents the cut between emitting and
  non-emitting galaxies. The BCGs for Abell 780 and Abell 1795 have unreliable H$\beta$ equivalent widths, but nonetheless strong emission, and are represented by lower limits at W$_{o}$~$>3.3$~\AA~and~$>3.7$~\AA, respectively. \underline{Bottom Panel}: Using the same
  subsamples as the top panel, we plot the fraction of line emitting
  galaxies as a function of K-band magnitude. The solid line represents the
  BCGs, and the dotted line the
  control galaxies. \label{relkcont}\label{hbkcont}}   
\end{figure*}

We show in Fig. \ref{hbkcont} the line emission strength as a function
of K absolute magnitude for the NFPS and SDSS BCGs and control
galaxies. The NFPS clusters are separated into CF and non-CF cases. 
In the non-CF clusters, there are no emitting BCGs, and the fraction of emission-line galaxies, shown
in the bottom panels, is also very low for the control galaxies ($\sim 5$~\%).  On the other hand, in CF clusters $\sim 70$~\% of the BCGs show emission, as we noticed also in the
previous section.  Moreover, the BCGs with strong emission tend to be
the brightest galaxies, $M_K<-25.5$, where by definition, there are
few control galaxies.  Hence, this trend is quite different from what
is seen in the control sample, where almost all of the emission line
galaxies are fainter than $M_K=-25$.  For magnitudes near $M_K\approx
-25$, where there is substantial overlap between the two populations,
the emission line fraction of both populations are similar for non-CF
clusters.  The results from the SDSS, which include all clusters
regardless of X-ray luminosity or CF status, are consistent with the
results for the total NFPS.  There are somewhat fewer emission-line BCGs at a given
magnitude than the for NFPS, presumably because CF clusters make up a
smaller proportion of the sample in the SDSS.  We will say more about
this in Section~\ref{discussion}. 

Also highlighted on Fig.~\ref{hbkcont} and Fig.~\ref{hbrcont} are the
BCGs with LINER-like emission. Most of the emitting BCGs are more
characteristic of LINER emission; this is especially true of the less luminous BCGs. 

\subsection{Dependence on Location in Cluster}\label{sec-loc}
In our samples, there are some BCGs that are found close to the X-ray
centre, while others can be found several hundred kpc away.  In this
section we will investigate whether or not the presence of emission
lines in the BCG depends on its distance to the X-ray centre. 

\begin{figure*}
\centering
\epsfysize=9cm
\epsfbox{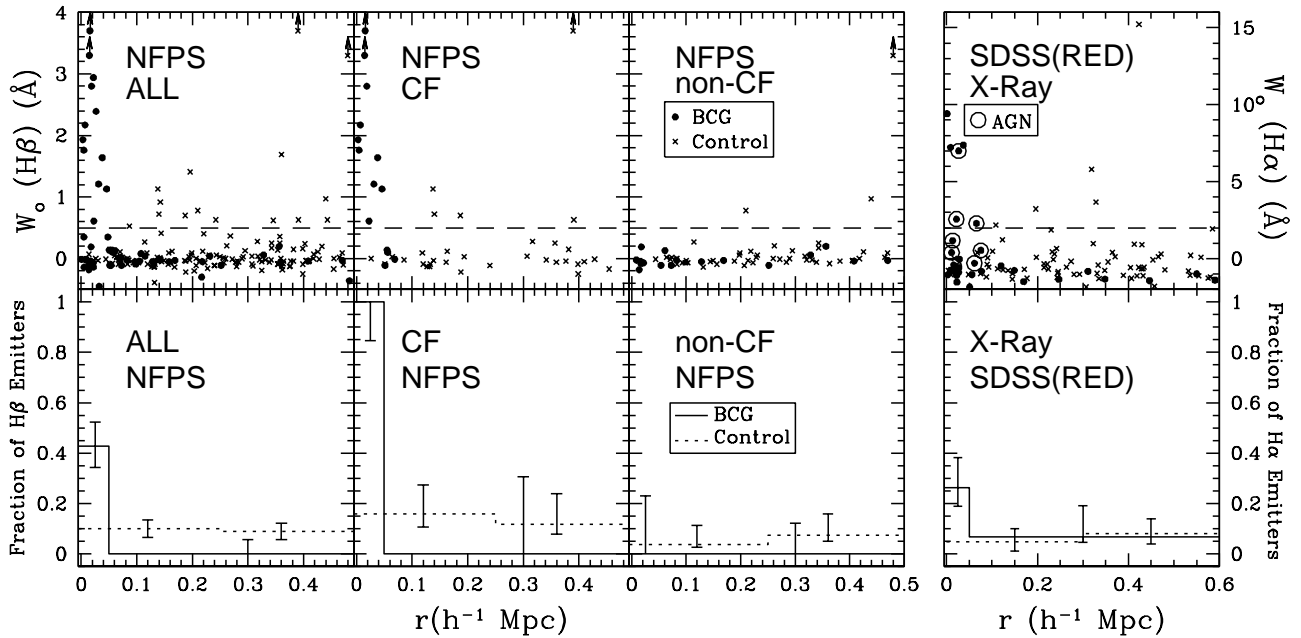}
\caption[Emission line strength as a function of Radius]{Line Emission
  as a function of distance from the cluster X-ray
  centre. Subsample definitions and symbols are the same as in Fig.~\ref{relkcont}. 
\label{hbrcont}}  
\end{figure*}

This is illustrated in Fig.~\ref{hbrcont} where we show the line
emission as a function of cluster radius for the NFPS galaxies. {\it
  All of the strongly emitting BCGs are within 50 kpc of the X-ray
  centre}.  As discussed in the previous section, these emitting
galaxies are also usually found in a cooling flow cluster.  In the
rightmost panel we show the equivalent plot for the 34 BCGs in the
SDSS sample for which we have X-ray centres.  Although the sample is
small, our results are consistent with those seen in the NFPS; only
those BCGs that are close to the X-ray centre have significant line
emission. 

A Kolmogorov-Smirnov test on the H$\beta$ distributions of the non-CF
BCGs and the controls shows no evidence for a difference in the two
populations. Therefore, to summarize, we conclude that an increased
frequency of optical line emission in BCGs is observed only for those
galaxies that lie within 50kpc of the X-ray centre of a CF cluster. 
For the remainder, the frequency of emission in BCGs is consistent
with that observed in the control population. 

\subsection{Dependence on Cluster Mass and Density}\label{sdss}

The SDSS provides us with an optically-selected sample, spanning a
wide range in velocity dispersion, that should be representative of
the cluster population independent of X-ray properties.  In this
section we use this full sample (including those galaxies with
($u'-r'$) $<2.2$ that were excluded when comparing directly with the
NFPS) to explore the effect of environment on the presence of line
emission in BCGs. 

\begin{figure*}
\epsfysize=3in
\epsfbox{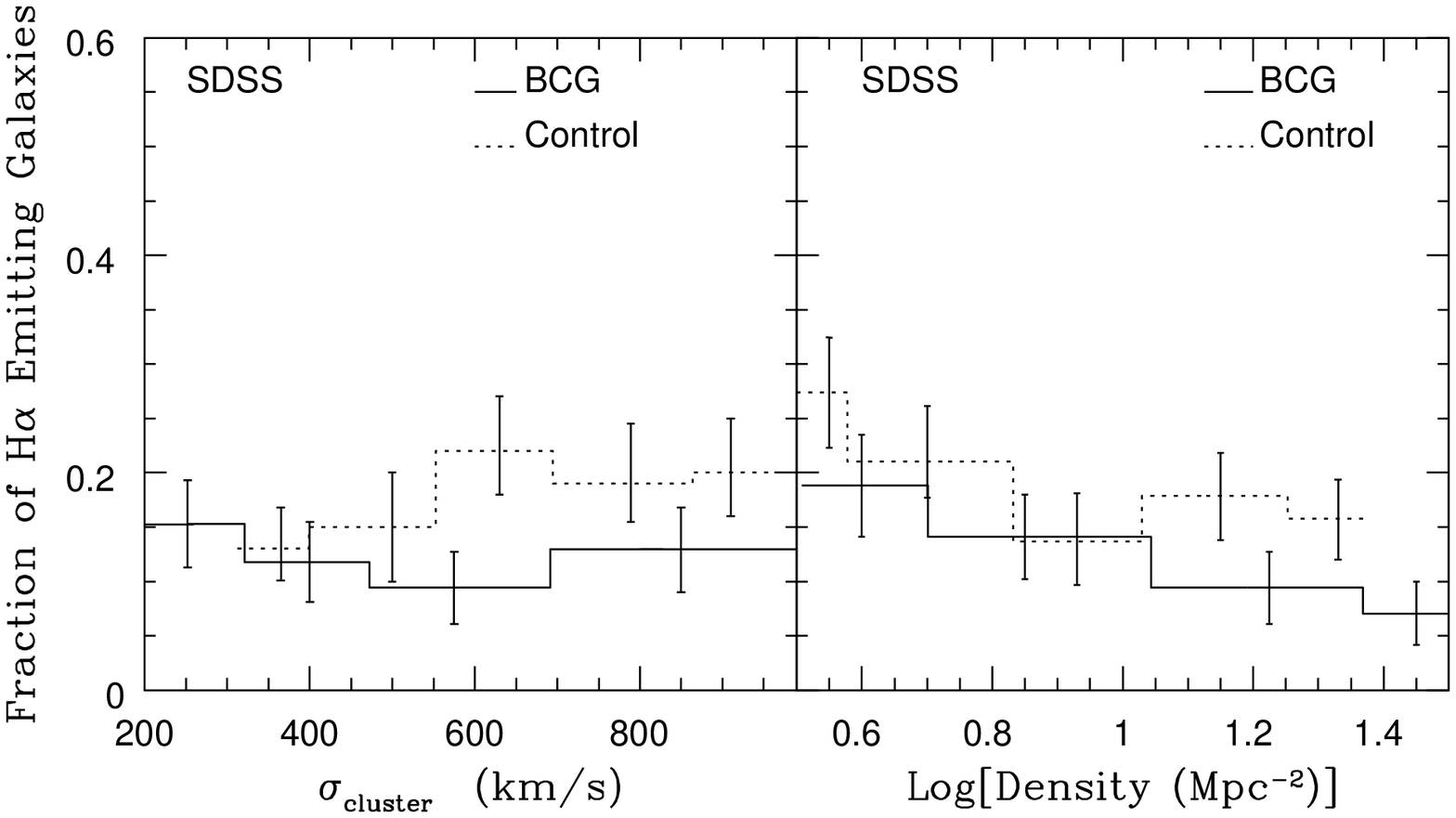}
\caption{Left Panel: The fraction of H$\alpha$ emitting
  galaxies in the SDSS as a function of group velocity dispersion. The
  solid line refers to BCGs and the dotted line are the control
  galaxies. This adaptive histogram contains 85 galaxies per bin for
  the BCGs and 100 galaxies per bin for the controls. Right Panel:  The fraction of H$\alpha$ emitting galaxies in the SDSS as a
  function of the galaxy density. This adaptive histogram contains 85 galaxies per
  bin for the BCGs and 95 galaxies per bin for the controls. 
\label{denshistb}
\label{adapsighist}
}
\end{figure*}

Since the cooling flow status of a cluster might be correlated with
its total mass, or central mass density, we wish to explore whether
the trends we have observed are merely reflecting a more fundamental
correlation with either of these quantities. First, in Fig. 
\ref{adapsighist}, we plot the fraction of BCGs with H$\alpha$
emission as a function of the cluster velocity dispersion.  There is
no strong trend. Rather, a fraction $\sim $~10-15~\% of BCGs in each bin showing
H$\alpha$ emission is observed (the Spearman correlation coefficient is 0.17).  We note that the frequency of radio-loud AGN is also found to be
independent of velocity dispersion \citep{bes06}. The
control galaxy population in the SDSS sample also shows no strong correlation
 between emission line fraction and group velocity
dispersion (the Spearman correlation coefficient is 0.38). 
For the control galaxies, the overall fraction of emitting
galaxies is somewhat higher than for BCGs, closer to 15-20\%. 

Next, we examine in Fig.~\ref{denshistb} the frequency of emission
lines in BCGs as a function of the galaxy density, as measured
by the distance to the fifth-nearest spectroscopic neighbour.  Even in the densest
regions this corresponds to a smoothing scale of $\sim200$ kpc, so the
measurement here is of the total mass density on scales larger than
the cooling radius.  Generally, one would expect the
central regions of CF clusters to be the densest environments, and if
the presence of a cooling-flow was correlated with the mass or galaxy
density on larger scales our previous results would lead us to expect
the most star formation occurring in the most 
dense regions. On the contrary, we observe a clear correlation with
the emission line fraction 
decreasing with increasing number density, and so BCGs are {\it less} likely to
show emission lines if they are found in the densest regions of clusters (in both cases of the BCG and the controls, the Spearman test yields a
correlation coefficient of $\sim 0.9$, and the correlation is significant at the $\sim 95$\% confidence level).  At all densities, the
fraction of emission-line galaxies is higher in the control population
than the BCG population.  The fact that we
observe enhanced emission for those galaxies within the smaller 50 kpc scale of the X-ray centre
of CF clusters (Section \ref{sec-loc}) is therefore very likely related directly to the
presence of cooling, X-ray gas on small scales, rather than the overall gravitational potential. 

\section{Discussion}\label{discussion}

The overall fraction of BCGs with emission lines is $\sim 13$~\% in
the SDSS, and $\sim 20$~\% in the NFPS.  The latter is in good
agreement with \citet{cra99} who find a fraction of 27~\%
emission-line BCGs in their sample of X-ray selected clusters. We do
not know what the CF fraction is in an optically-selected sample, or
as a function of mass. Nonetheless, since 
the fraction of massive clusters that host a cooling flow is likely no
more than about 50 \% 
\citep{per98,mcc04,che07}, and CF clusters are systematically
overluminous, we attribute the factor of two difference in
emission fraction between our two surveys to the fact that NFPS
\citep[and][]{cra99} is
X-ray selected, and therefore biased toward CF clusters. 

There is currently much observational and theoretical work
exploring the possible feedback between a central galaxy's AGN,
current star formation, and the cooling X-ray gas
\citep[e.g.]{sil98,bes06,cro06,bow06,sij06,del06}.  AGN in the BCGs
are thought to play an important role in suppressing cooling, and
lowering mass deposition rates (and hence star formation rates).  Our
main result is that the majority of line emitting BCGs are positioned
close to the X-ray centre of clusters classified as hosting a cooling
flow directly from the mass deposition rates (as defined in Section \ref{nfpss}). Furthermore, this result holds when we classify cooling flow clusters instead by their excess
X-ray luminosity relative to other clusters with similar dynamical
mass (a definition we explored at the end of Section \ref{sec-cfdef}).  Similar conclusions, based on smaller samples, have been
reached by others \citep[e.g.][]{joh87,mcn89,cra99,raf06}. 
Importantly, we have shown that, in the absence of a cooling flow,
emission lines are rare in BCGs, regardless of the mass density or
velocity dispersion of the cluster.  Moreover, a control population of
similarly bright, centrally located cluster galaxies does not exhibit
an increased frequency of emission lines in CF clusters.  We therefore
conclude that the observed emission (arising from star formation
and/or an AGN) is directly related to the presence of cooling gas. 

For the most part, we have not concerned ourselves with the origin of
the observed emission, since starburst galaxies and optically-selected
AGNs are probably closely linked \citep{kau03}. However, it is worth
investigating this further here.   \citet{cra99} find that
the very strongest H$\alpha$ emitters have star formation-like
emission line ratios, while \citet{von06} find that most BCGs, which display weaker H$\alpha$
emission, are more characteristic of LINER activity. The six NFPS BCGs that have [NII]/H$\alpha$ line ratios available from
\citet{cra99} lie in the regime straddling LINERS and Seyferts. Of the
six emitting BCGs in the SDSS-X-Ray sample, the three stronger
H$\alpha$ emitters are non-AGN, which means that they are likely
composite, whereas the somewhat weaker emitters are
classified as AGN. Thus the emission-line BCGS found near the X-ray
centre of galaxy clusters appear to be a
heterogeneous class of objects.

In the unrestricted SDSS sample, we find only $\sim 13$~\% of BCGs
have line emission, compared to $\sim 18$~\% for controls. This
exhibits the same trend as in the \citet{von06} study of SDSS C4
cluster galaxies, where detectable emission line luminosity
measurements (with S/N $>$3) are found in 30~\% of BCGs, as opposed to
40~\% for the other massive, central, galaxies (ie. controls). Our
numbers cannot be compared directly, as their cut includes more weakly
emitting systems as emitters than does our more strict criterion of
W$_{o}$~$> 2$  \AA. If we relax our definition of an emitting galaxy being one with W$_{o}$~$> 0$ \AA, we find the percentage of line emitters increases to 27~\% for BCGs, and 38~\% for controls.

Our Fig.~\ref{denshistb} reiterates the above point, that in the SDSS sample,
the BCGs have a lower fraction of line emitting galaxies than the
controls. This is also consistent with what \citet{bes06} find: that
emission line AGN activity is suppressed in the galaxies near the
centre of the cluster with respect to the other massive cluster
galaxies.  They point out that emission line AGN and radio-loud AGN
are partly independent populations, but that BCGs with emission-line
AGN are more likely to host radio-loud AGN than other galaxies. Best
et al. also find that radio-loud AGN are preferentially found in BCGs
within 0.2 r$_{200}$ of the centre of the cluster and that there is a
strong dependence on galaxy stellar mass, with $\sim 40$~\% of the
most massive galaxies showing radio loud AGN emission. Therefore it is
plausible that our emission-line BCGs in the NFPS (which are found at
the centre of CF clusters) and in the SDSS (many of which show
LINER-like activity), host radio-loud AGN. 
 We therefore matched our NFPS BCGs to radio sources from the Faint
 Images of the Radio Sky at Twenty-Centimeters survey \citep[hereafter FIRST]{wbh97} and NRAO
 VLA Sky Survey \citep[hereafter NVSS]{con98}.  We find that indeed, all 12 of our H$\beta$
 emitting BCGs are radio sources. Furthermore, for the 50 BCGs in the
 FOV of the surveys, 28 of our BCGs have radio counterparts
 ($\sim 56$\ensuremath{\pm11}~\%). Thus~$\sim42$\ensuremath{\pm11}~\% of non-H$\beta$ emitting BCGs have radio emission. Again, the cooling flow status of the cluster appears important:~$\sim71^{+~9}_{-14}$ \% of CF BCGs are radio sources, and~$\sim80^{+~7}_{-16}$ \% of central CF BCGs are radio sources. For non-CF BCGs only~$\sim32^{+12}_{-~9}$ \% have radio emission. For the controls, 36 of 144 galaxies in the FOV of the surveys are radio sources ($\sim 25$\ensuremath{\pm4}~\%). 
  
Recently, \citet{lin06} have studied the radio-loud properties of a large
X-ray selected cluster sample, and find the overall radio-loud fraction
to be $\sim 35$~\% for K-band selected BCGs within r$_{200}$ of the cluster
center (compared with $\sim 20$~\% for bright cluster galaxies, excluding
the BCGs).   This is in reasonable agreement with our results, although
the radio-loud BCG fraction we measure is somewhat
higher. \citet{lin06} also find a strong trend with 
cluster mass (inferred from the X-ray luminosity), though the strength
of the trend is sensitive to the radio power limit and 
K-band luminosity of the galaxy. Our NFPS sample is too small to
robustly identify such trends, but we note that the fraction of BCGs
with radio sources that are in clusters with a velocity dispersion $<$ 600km/s is
54\ensuremath{\pm15}~\%, quite similar to that for BCGs in clusters with
a velocity dispersion $>$ 600km/s, 58\ensuremath{\pm15}~\%.  In the
optically-selected SDSS, we find no significant trend with velocity
dispersion, but an overall fraction of~$\sim 40$~\% of the BCGs with radio emission, and~$\sim 25$~\% of the controls.  Because of the difference in sample selection, and the
cluster mass estimators, we do not consider this a
serious discrepancy with the results of \citet{lin06}.

\section{conclusions}\label{conclusion}

We have used two large, homogeneous galaxy cluster surveys to
investigate the incidence of optical emission lines amongst BCGs.  The
NFPS consists of 60 BCGs in X-ray selected clusters, while the SDSS
sample is a larger, optically-selected sample of 328 BCGs.  From these
data, we are able to draw the following conclusions:
\begin{itemize}
\item Of the 10 BCGs that lie within 50kpc of the peak of X-ray
  emission in a cluster with evidence for a significant cooling flow,
  all show optical emission lines. Moreover, of the 12 BCGs
  that show emission, all are located within 50kpc of the X-ray
  emission peak, all are radio sources, and none are in clusters of known non-CF status (10 are in CF clusters, and 2 are in clusters with unknown CF status). 
\item Excluding the special circumstances noted above, the fraction of
  BCGs that exhibit optical emission lines is $\sim $~10-20 \%, and is
  always comparable to or lower than the fraction for control galaxies
  with a similar luminosity and environment. 
\item For optically selected cluster samples, which are dominated by
  non-CF clusters, the fraction of BCGs with emission does not
  correlate strongly with cluster mass or galaxy density. 
\end{itemize}

We have therefore demonstrated a direct connection between the
presence of cooling gas, and enhanced optical emission in a centrally
located galaxy.  It would be very useful to obtain pointed X-ray
observations of those SDSS clusters in which we have found a BCG with
H$\alpha$ emission, to determine if this correlation holds in
optically-selected samples.  These clusters would also be potentially
interesting for observation with {\it Chandra} to observe the X-ray emission morphology, as other massive clusters
with H$\alpha$ emitting BCGs, such as Abell 426 and Abell 1795 show
remarkable X-ray morphology, such as X-ray holes, and cooling tails. 

\begin{table*}

    \caption{Table of NFPS BCGs. The cluster name is shown in column 1. Column 2 and 3 give the position of the cluster X-ray centre. The cluster redshift is given in column 4, the cluster velocity dispersion and r$_{200}$ are given in columns 5 and 6. Column 7 gives the cooling flow mass deposition rate (MDR) in (M$_{\odot}$/yr).  Column 8 gives the cooling flow status of the cluster. Column 9 gives the reference for the MDR/CF status:  b stands for \citet{bir04}, f for \citet{fuj06}, j for \citet{joh05}, h for \citet{hen04}, k for \citet{kem04}, m for \citet{mcc04}, p for \citet{per98},  r for \citet{sha04}, s for \citet{san06}, v for \citet{kan06},  w for \citet{whi00}, and o for other. Column 10 is the X-ray luminosity in units of 10$^{44}$erg/s, column 11 gives the name of the BCG, column 12 is the BCG K-band magnitude, column 13 gives the distance between the BCG and the cluster X-ray centre. Columns 14 gives the H$\beta$ equivalent width, and column 15 gives the error, which is the noise in the line-free regions of the absorption-corrected spectrum.}
\scriptsize
 \label{bcgtab}
 \begin{tabular}{lccccccccclcccc} \hline \hline

Name & RA$_{x}$ & DEC$_{x}$ & z & $\sigma_{cl}$ & r$_{200}$ & MDR& CF &ref & L$_{X}$ & Name & M$_{K}$ & dr & $W_{o}$H$\beta$ & H$\beta$$_{err}$  \\ 
(Clus) & (deg) & (deg) &  & (km/s) & (Mpc) & (M$_{\odot}$/yr)  & &  & & (BCG) & (BCG) & (Mpc) & (\AA) & (\AA) \\
\hline

A0085 & 10.5 & -9.3 & 0.0557 & 736 & 1.3 & 108 & \checkmark & p/s & 4.920 & MCG--02-02-086 & -26.0 & 0.046 & 1.13 & 0.21 \\
A0119 & 14.1 & -1.2 & 0.0436 & 653 & 1.1 & 0 & X& p & 1.580 & UGC-00579 & -25.7 & 0.054 & -0.11 & 0.06 \\
A0133 & 15.7 & -21.9 & 0.0561 & 794 & 1.4 & 25 & X& b/m & 1.590 & ESO-541--G-013 & -25.6 & 0.017 & -0.08 & 0.09 \\
A0262 & 28.2 & 36.2 & 0.0155 & 432 & 0.7 & 2 & \checkmark& b/s & 0.230 & NGC-0708 & -24.8 & 0.005 & 1.76 & 0.06 \\
A0376 & 41.5 & 36.9 & 0.0482 & 975 & 1.7 & 42 & X & w & 0.680 & GIN-138 & -24.4 & 0.408 &-0.04 & 0.06 \\
A0407 & 45.5 & 35.8 & 0.0465 & 670 & 1.2 & N/A & ?& ?& 0.210 & UGC-02489-NED02 & -25.6 & 0.022 &-0.14 & 0.10 \\
A3128 & 52.6 & -52.5 & 0.0595 & 838 & 1.5 & N/A & ?& ?& 1.160 & 2MASX-J03295060-5234471 & -25.5 & 0.217 &-0.30 & 0.13 \\
RXJ0341 & 55.3 & 15.4 & 0.0288 & 502 & 0.9 & N/A & ?& ?& 0.250 & 2MASX-J03412829+1515326 & -24.5 & 0.231 & 0.04 & 0.11 \\
A3158 & 55.7 & -53.6 & 0.0586 & 814 & 1.4 & 292& \checkmark & w & 2.820 & ESO-156--G-008-NED01 & -25.7 & 0.069 &-0.01 & 0.09\\
A3266 & 67.9 & -61.4 & 0.0588 & 946 & 1.6 & 145 & X& w/m & 3.900 & ESO-118-IG-030-NED02 & -26.2 & 0.128 &-0.05 & 0.15 \\
A0496 & 68.4 & -13.2 & 0.0321 & 577 & 1.0 & 70 & \checkmark& o/m & 1.770 & MCG--02-12-039 & -25.5 & 0.031 & 1.21 & 0.18 \\
A3341 & 81.4 & -31.6 & 0.0376 & 500 & 0.9 & N/A & ?& ?& 0.310 & MCG--05-13-019 & -24.9 & 0.009 &-0.01 & 0.07 \\
A0548A & 87.2 & -25.4 & 0.0386 & 794 & 1.4 & 10 & X& w & 0.370 & ESO-488-IG-031 & -24.3 & 0.356 & 0.20 & 0.13 \\
A3376 & 90.4 & -40.0 & 0.0464 & 710 & 1.2 & 0 & X& w & 1.320 & ESO-307--G-013 & -25.3 & 0.470 &-0.03 & 0.06 \\
A3389 & 95.5 & -65.0 & 0.0270 & 626 & 1.1 & 25& X & w & 0.180 & NGC-2235 & -25.2 & 0.061 & 0.13 & 0.07 \\
A3391 & 96.6 & -53.7 & 0.0556 & 696 & 1.2 & 131& \checkmark & w & 1.370 & ESO-161-IG-007-NED02 & -26.4 & 0.055 & 0.14 & 0.10 \\
A3395 & 96.8 & -54.5 & 0.0491 & 640 & 1.1 & N/A & ?& ?& 1.610 & ESO-161--G-008 & -25.4 & 0.158 &-0.05 & 0.12 \\
A0576 & 110.4 & 55.8 & 0.0383 & 854 & 1.5 & 3 & X& p/k & 0.710 & CGCG-261-056-NED01 & -25.9 & 0.008 &-0.02 & 0.05 \\
UGC03957 & 115.2 & 55.4 & 0.0339 & 511 & 0.9 & N/A& ? & ?& 0.480 & UGC-03957 & -25.3 & 0.017 &-0.04 & 0.05 \\
A0602 & 118.4 & 29.4 & 0.0605 & 675 & 1.2 & N/A & ?& ?& 0.530 & 2MASX-J07532661+2921341 & -24.2 & 0.032 &-0.45 & 0.12 \\
Z1665 & 125.8 & 4.4 & 0.0296 & 437 & 0.8 & N/A & ?& ?& 0.160 & IC-0505 & -24.9 & 0.072 & 0.02 & 0.05 \\
A0754 & 137.3 & -9.7 & 0.0546 & 784 & 1.4 & 218& X & w/h & 4.460 & 2MASX-J09083238-0937470 & -25.7 & 0.328 &0.06 & 0.10 \\
A0757 & 138.4 & 47.7 & 0.0514 & 381 & 0.7 & N/A & ?& ?& 0.460 & 2MASX-J09134460+4742169 & -24.5 & 0.142 &-0.01 & 0.10 \\
A0780 & 139.5 & -12.1 & 0.0551 & 641 & 1.1 & 492 & \checkmark& w/m & 3.470 & Hydra-A & -25.1 & 0.015 &  7.8 & 0.30 \\
Z2844 & 150.7 & 32.7 & 0.0504 & 462 & 0.8 & N/A & ?& ?& 0.300 & NGC-3099 & -25.4 & 0.048 & 0.35 & 0.07 \\
A1367 & 176.2 & 19.8 & 0.0219 & 747 & 1.3 & 0 & X& o/m & 0.930 & NGC-3842 & -24.9 & 0.252 &-0.11 & 0.05 \\
Z4803 & 181.1 & 1.9 & 0.0206 & 474 & 0.8 & N/A & ?& ?& 0.170 & NGC-4073 & -25.4 & 0.000 &-0.01 & 0.05 \\
A1631A & 193.2 & -15.4 & 0.0461 & 531 & 0.9 & N/A & ?&?& 0.330 & 2MASX-J12523166-1512150 & -24.4 & 0.358 &-0.06 & 0.07 \\
A3528B & 193.6 & -29.0 & 0.0547 & 500 & 0.9 & N/A & ?& ?& 0.690 & ESO-443--G-004 & -25.8 & 0.005 &-0.14 & 0.07 \\
A3528A & 193.7 & -29.3 & 0.0535 & 698 & 1.2 & N/A & ?& ?& 1.100 & 2MASX-J12543999-2927327 & -24.0 & 0.482 &-0.36 & 0.25 \\
A3530 & 193.9 & -30.4 & 0.0544 & 436 & 0.8 & N/A & ?& ?& 0.380 & AM-1252-300-NED02 & -25.0 & 0.018 &N/A & N/A \\
A1656 & 194.9 & 27.9 & 0.0230 & 898 & 1.6 & 85 & X& w & 3.980 & NGC-4889 & -25.6 & 0.169 &-0.03 & 0.03 \\
A1668 & 195.9 & 19.3 & 0.0641 & 476 & 0.8 & N/A & ?& ?& 0.880 & IC-4130 & -25.2 & 0.027 & 2.39 & 0.09 \\
A3558 & 202.0 & -31.5 & 0.0476 & 814 & 1.4 & 235 & X& w/s & 3.450 & ESO-444--G-046 & -26.2 & 0.019 &-0.06 & 0.12 \\
A1736A & 201.7 & -27.1 & 0.0465 & 664 & 1.2 & 79 & X& w & 1.250 & IC-4252 & -25.8 & 0.512 &-0.02 & 0.04 \\
A3560 & 203.1 & -33.1 & 0.0487 & 548 & 0.9 & N/A & ?& ?& 0.730 & 2MASX-J13322574-3308100 & -25.2 & 0.097 &-0.05 & 0.06 \\
A3571 & 206.9 & -32.9 & 0.0392 & 913 & 1.6 & 130& X & w/s & 3.910 & ESO-383--G-076 & -26.0 & 0.022 &-0.07 & 0.14 \\
A1795 & 207.2 & 26.6 & 0.0627 & 725 & 1.3 & 18 & \checkmark& b/m & 6.330 & CGCG-162-010 & -25.7 & 0.016 & 7.2 & 0.30 \\
A3581 & 211.9 & -27.0 & 0.0225 & 525 & 0.9 & 18 & \checkmark& w/j & 0.320 & IC-4374 & -24.5 & 0.003 & 1.93 & 0.09 \\
A1983A & 223.2 & 16.7 & 0.0448 & 472 & 0.8 & N/A & ?& ?& 0.230 & ABELL-1983-1:[CBW93]-C & -24.2 & 0.050 & 0.14 & 0.09 \\
A1991 & 223.6 & 18.6 & 0.0589 & 454 & 0.8 & 37 & \checkmark& w/r & 0.710 & NGC-5778 & -25.4 & 0.023 & 0.61 & 0.10\\
A2052 & 229.2 & 7.0 & 0.0352 & 531 & 0.9 & 81 & \checkmark&b/m & 1.330 & UGC-09799 & -25.5 & 0.038 & 1.64 & 0.09 \\
A2063 & 230.8 & 8.6 & 0.0344 & 764 & 1.3 & 99 & \checkmark& w/v & 1.020 & CGCG-077-097 & -24.9 & 0.056 & 0.10 & 0.08 \\
A2107 & 234.9 & 21.8 & 0.0415 & 527 & 0.9 & 57& X & w/f & 0.570 & UGC-09958 & -25.5 & 0.014 &-0.18 & 0.06 \\
A2147 & 240.6 & 16.0 & 0.0370 & 711 & 1.2 & 119& X & w/s & 1.660 & UGC-10143 & -24.9 & 0.082 &-0.02 & 0.07 \\
A2151A & 241.2 & 17.7 & 0.0352 & 746 & 1.3 & 173 & \checkmark& w & 0.430 & NGC-6041A & -25.1 & 0.051 &-0.11 & 0.07 \\
A2199 & 247.2 & 39.5 & 0.0293 & 647 & 1.1 & 2 & \checkmark&b/m & 1.900 & NGC-6166-NED01 & -25.7 & 0.007 & 2.17 & 0.11 \\
RXJ1733 & 263.3 & 43.8 & 0.0319 & 468 & 0.8 & N/A & ?& ?& 0.270 & IC-1262 & -24.6 & 0.005 & 0.35 & 0.08 \\
RXJ1740 & 265.1 & 35.7 & 0.0434 & 556 & 1.0 & N/A & ?& ?& 0.270 & CGCG-199-007-NED01 & -24.9 & 0.013 &-0.12 & 0.08 \\
Z8338 & 272.7 & 49.9 & 0.0494 & 532 & 0.9 & N/A & ?& ?& 0.450 & NGC-6582-NED02 & -25.7 & 0.098 &-0.09 & 0.04 \\
A3667 & 303.1 & -56.8 & 0.0549 & 852 & 1.5 & 196 & X& w/m & 5.020 & IC-4965 & -25.8 & 0.700 &N/A & N/A \\
A3716 & 312.9 & -52.7 & 0.0447 & 748 & 1.3 & N/A & ?& ?& 0.480 & ESO-187--G-020 & -25.2 & 0.098 &-0.07 & 0.08 \\
IIZW108 & 318.5 & 2.6 & 0.0482 & 399 & 0.7 & N/A & ?& ?& 1.090 & IC-1365-NED02 & -25.6 & 0.115 & 0.04 & 0.06 \\
A2399 & 329.3 & -7.8 & 0.0577 & 608 & 1.1 & N/A & ?& ?& 0.430 & 2MASX-J21572939-0747443 & -24.5 & 0.218 &-0.05 & 0.06 \\
A3880 & 337.0 & -30.6 & 0.0583 & 817 & 1.4 & N/A & ?& ?& 0.960 & PKS-2225-308 & -25.6 & 0.022 & 2.94 & 0.15 \\
Z8852 & 347.6 & 7.6 & 0.0402 & 771 & 1.3 & N/A & ?& ?& 0.470 & NGC-7503 & -25.6 & 0.108 & 0.08 & 0.07 \\
A2572B & 349.3 & 18.7 & 0.0386 & 138 & 0.2 & N/A & ?& ?& 0.280 & NGC-7602 & -24.5 & 0.105 &-0.04 & 0.05 \\
A2589 & 351.0 & 16.8 & 0.0411 & 583 & 1.0 & 0& X &o/m & 0.890 & NGC-7647 & -25.4 & 0.073 &-0.11 & 0.07 \\
A2634 & 354.6 & 27.0 & 0.0309 & 692 & 1.2 & 0 & X& w & 0.480 & NGC-7720-NED01 & -25.5 & 0.018 & 0.19 & 0.08\\
A4059 & 359.2 & -34.8 & 0.0496 & 556 & 1.0 & 7 & \checkmark& b & 1.680 & ESO-349--G-010 & -26.0 & 0.019 & 2.80 & 0.12 \\
\end{tabular}

\end{table*}


\section*{Acknowledgments}
We are very grateful to C. Miller and R. Nichol for their help with
the SDSS database, and the C4 cluster catalogue.  We also thank R. 
Finn for useful discussions about the C4 clusters.  LOVE wishes to
thank C. Robert for helpful comments and support during her stay at
the University of Waterloo. MJH and MLB acknowledge support from their
respective NSERC Discovery grants.  We also thank the anonymous referee for a careful reading of the manuscript and 
useful suggestions.
\bibliography{le1}
\end{document}